\documentclass[letterpaper, 10 pt, conference]{ieeeconf} 
\IEEEoverridecommandlockouts
\usepackage{tikz}
\usetikzlibrary{tikzmark, positioning, fit, shapes.misc}
\usetikzlibrary{decorations.pathreplacing, calc}
\usepackage{authblk}

\usepackage{cite}
\usepackage{amsmath,amssymb,amsfonts}
\usepackage{algorithmic}
\usepackage{graphicx}
\usepackage{textcomp}
\usepackage{array}
\usepackage{xcolor}
\usepackage{eurosym}
\usepackage[hidelinks]{hyperref}
\usepackage{booktabs,floatrow}
\DeclareMathAlphabet{\mathpzc}{T1}{pzc}{m}{it}

\DeclareMathOperator*{\argmin}{arg\,min}
\newcommand{\suchthat}{\;\ifnum\currentgrouptype=16 \middle\fi|\;}

\usepackage{etoolbox}
\makeatletter
\patchcmd{\@makecaption}
  {\scshape}
  {}
  {}
  {}
\makeatother

\begin{document}
\author[1]{Mirhan Ürkmez\thanks{This work is funded by Independent Research Fund Denmark (DFF). We thank Verdo, Peter Nordahn, and Steffen Schmidt for providing the EPANET model(An open source hydraulic simulator widely used by utilities\cite{epaEPANET}.).}}
\author[1,2]{Carsten Kallesøe}
\author[1]{Jan Dimon Bendtsen}
\author[3,4]{Eric C. Kerrigan}
\author[1]{John Leth}
\affil[1]{\textit{Aalborg University, Fredrik Bajers Vej 7c, DK-9220 Aalborg,
Denmark} \protect}
\affil[2]{\textit{Grundfos Holdings A/S, Bjerringbro, Denmark}}
\affil[3]{\textit{Department of Electrical and Electronic Engineering, Imperial
College London, UK} \protect}
\affil[4]{\textit{Department of Aeronautics, Imperial College London, UK} \protect \\
(e-mail: \{mu,csk,dimon,jjl\}@es.aau.dk, e.kerrigan@imperial.ac.uk)}

\title{A Robust Predictive Control Method  for Pump Scheduling in Water Distribution Networks} 

\maketitle

\begin{abstract}
Water utilities aim to reduce the high electrical costs of Water Distribution Networks (WDNs), primarily driven by pumping. However, pump scheduling is challenging due to model uncertainties and water demand forecast errors. This paper presents a Robust Model Predictive Control (RMPC) method for optimal and reliable pump scheduling, extending a previous efficient robust control method tailored to our model.  A linear model with bounded additive disturbances is used to represent tank water level evolution, with uncertainty bounds derived from WDN simulation and demand data. At each time step, a pump scheduling policy, affine in past disturbances, is optimized to satisfy system constraints over a prediction horizon. The resulting policies are then applied in a receding horizon fashion. The optimization problem is formulated to require $\mathcal{O}(N^6)$ computations per iteration with an interior-point method, which is reduced to $\mathcal{O}(N^3)$ by reformulating it into a sparse form.  When evaluated on a model representing the water distribution network of Randers, a medium-sized town in Denmark, the method surpasses nominal and constraint-tightening model predictive control (MPC) approaches in terms of meeting constraints and provides comparable economic outcomes.
\end{abstract}


\section{Introduction}

Water systems account for 7-8\% of global electricity use, mainly through pumping \cite{WSSEnergy}.  To reduce electricity costs, researchers focused on optimizing Water Distribution Network (WDN) pump schedules \cite{MALAJETMAROVA2017209}. However, the nonlinear nature and large scale of WDNs make scheduling challenging, often requiring simplified models for optimization \cite{Urkmez2024OptimizingPP, Wang2018EconomicMP}. These simplified representations introduce uncertainties from model-plant mismatch, water demand predictions, and network parameter variations. While most existing work overlooks these uncertainties, addressing them is essential for ensuring reliable water service

Several studies in the literature adress uncertainties in pump scheduling using stochastic methods. Chance constraints have been applied to ensure demand satisfaction \cite{chance1} and maintain tank levels \cite{grosso}. Scenario-based approaches have also been explored, such as a tree-based scenario method in \cite{PALLOTTINO20051031} and two stochastic MPC formulations in \cite{grosso}, one with a uniform control sequence and another with distinct sequences per scenario. More recently, a distributionally robust approach \cite{9765702} is used, where instead of assuming a fixed probability distribution, the probability distribution is assumed to be in an ambiguity set.

Several studies address uncertainties using bounded sets rather than probability distributions for robust WDN operation. In \cite{Goryashko2014RobustEC} and \cite{perelman1}, input policies that are affine in demand are repeatedly calculated in a receding horizon fashion to ensure the robust satisfaction of the tank level constraints of WDNs. These studies assume a simplified WDN structure where each pump supplies water to one tank, allowing tank levels to be modeled exactly as a linear function of pump flows and water demands. This assumption removes model-plant mismatch uncertainty, so the focus is solely on uncertainty in water demands. Similarly, \cite{7844498} uses demand-affine policies with offline coefficients and applies tube-based MPC by tightening constraints based on these coefficients. Another constraint-tightening-based approach is used in \cite{WANG20175202}, where tightenings are calculated using interval arithmetic.


This work presents a robust MPC-based pump scheduling approach for WDNs with elevated tanks. Existing studies use either nonlinear models, which make MPC computationally expensive, or linear models that often neglect model uncertainty, affecting constraint satisfaction. We propose a surrogate model where tank dynamics are expressed as a linear function of the state, control input, and demand, with an additive disturbance term capturing both demand forecast and model uncertainties. The first contribution is the development of a method to quantify the uncertainty of this surrogate model in WDNs, enabling efficient computation within the robust MPC framework.   The second is the integration of a robust MPC method, adapting the formulation from \cite{Goulart2008EfficientRO} to accommodate the additional demand term with a per-iteration complexity of 
$\mathcal{O}(N^3)$ using primal-dual interior-point methods. We apply the proposed method to a linear model of the WDN in Randers, Denmark, and show that it outperforms several constraint-tightening MPC methods in constraint satisfaction while remaining cost-efficient.

\section{Water Distribution Network Model}
\label{sec:network}
In this section, a control-oriented model of WDNs is presented. A WDN consists of several types of components including pipes, demand nodes, pumps, valves, and tanks. Tanks are dynamic elements whose water levels change in time while the behavior of other elements can be represented with algebraic equations. We use these water levels as dynamic states and develop a data-based black box model to approximate their changes.  For brevity, we omit detailed explanations of other network elements such as pipes and nodes; for details on these components, see \cite{Urkmez2024OptimizingPP}.

The water tanks are dynamic elements with a water level changing with time according to 
\begin{align}\textstyle
 A^{j}\dot{{h}}^{j} = \sum_{i \in \mathcal{N}_j} q^{ij},
 \label{eq:tankLevel}
\end{align}
where $A^{j}$ represents the cross-sectional area of the tank identified by index $j$, $h^{j}$ is the tank level, $q^{ij}$ is the flow entering the tank, and $\mathcal{N}_j$ denotes the set of neighbor nodes of node $j$, viewing the WDN as a graph, where nodes represent tanks and junctions while edges correspond to pipes. These flows  $q^{ij}$ are nonlinear functions of the tank levels, pump flows, and the demand of each node in the network, due to nonlinear pipe and valve behaviors. However, as individual demand data is often unavailable, we assume only historical data of the aggregated demand for the pump-served area is accessible.  Using this aggregated demand, we approximate the flow rates of the pipes $q^{ij}$ connected to the tanks  with the following surrogate model
\begin{align}
    q^{ij}(t)= a^{ij}h(t)+b_1^{ij}u(t)+b_2^{ij}d_a(t)+\epsilon^{ij}(t),
    \label{eq:lspipe}
\end{align}
where  $h\in \mathbb{R}^{n},u \in \mathbb{R}^{m}$ are the vectors containing the tank levels and pump flow rates respectively, $d_{a}(t)$ is the aggregated demand at time $t$, $\epsilon^{ij}(t) \in \mathbb{R}$ is the residual, $a^{ij} \in \mathbb{R}^{1 \times n},b_1^{ij} \in \mathbb{R}^{1 \times m},b_2^{ij} \in \mathbb{R}$ are the constants to be estimated. To determine the values of these constants, the EPANET model of the WDN is first simulated with various initial conditions and pump flow rates to generate a dataset. Then, the values of the constants are found by a least squares fit to the data set.   Combining \eqref{eq:lspipe} with the tank level equations \eqref{eq:tankLevel} 
and putting the tank dynamics into vector form give
\begin{subequations}
    \begin{align}
    \dot{{h}}&= Ah(t) + B_1u(t) + B_2 d_{a}(t)+w_m(t), \label{eq:reducedModel} \\
    \label{eq:A-wm}
                        A &= \begin{bmatrix} \frac{1}{A_1}\sum_{i \in \mathcal{N}_1} a^{i1} \\ \vdots \\ \frac{1}{A_n}\sum_{i \in \mathcal{N}_n} a^{in}
                        \end{bmatrix},  w_m(t)= \begin{bmatrix} \frac{1}{A_1}\sum_{i \in \mathcal{N}_1} \epsilon^{i1}(t) \\ \vdots \\ \frac{1}{A_n}\sum_{i \in \mathcal{N}_n} \epsilon^{in}(t)
                         \end{bmatrix}.
    \end{align}
\end{subequations}
where $A \in \mathbb{R}^{n\times n}$, $B_1 \in \mathbb{R}^{n\times m}$, $B_2 \in \mathbb{R}^{n\times 1}$ are constant system matrices, and $w_m\in \mathbb{R}^n$ is the model error. The matrices $B_1$, $B_2$,  follow the same structure as $A$, with $a^{ij}$ replaced by $b_1^{ij}$, and $b_2^{ij}$ respectively. For a discussion on the suitability of this model for WDNs, please see \cite{Urkmez2024OptimizingPP}.

The inlet pressure, $p^{in}$, and the outlet pressure, $p^{out}$, of the pumps are required for calculating the electricity costs. It is assumed that the inlet pressure $p^{in}$ remains constant. The outlet pressure is modeled as a linear function of the states and inputs, which can be expressed as $p^{out}(t)=Ch(t)+Du(t)$,
where $C  \in \mathbb{R}^{m\times n}$ and $D \in \mathbb{R}^{m\times m}$ are identified from the data generated by the EPANET model. 

\subsection{Uncertainty Quantification}
In this work, the model errors $w_m(t)$ are assumed to be bounded and belong to a model uncertainty set $w_m(t) \in W_m=\{w\in \mathbb{R}^n\mid w=E_mg,\left|\left|g\right|\right|_\infty\leq 1\}$ where  $E_m \in \mathbb{R}^{n\times l}$. The set $W_m$ is identified empirically from EPANET simulation data.  Specifically, the model error values $w_m(t)$ are obtained using \eqref{eq:A-wm} and $E_m$ is determined such that the model uncertainty set $W_m$ includes all the model error  data $\{ w_m^{j} \}_{j=1}^{S}$ with $S \in \mathbb{R}$ samples. The matrix $E_m$ is set to 
\begin{equation}
    E_m=diag((w_m^{1})_{max},\cdots,(w_m^{n})_{max}).
\end{equation}
where $(w_m^{i})_{max}\in \mathbb{R}$ is the maximum value of the absolute value of the $i^{th}$ element of the model error $w_m$ in the simulation data i.e. $(w_m^{i})_{max}=\max_{j \in \{1, 2, \dots, S\}} \lvert w_m^{i,j}\rvert$, where $ w_m^{i,j}$ is the $i^{th}$ element of the $j^{th}$ sample of $w_m$ in the simulation data.

Another source of uncertainty comes from water demand predictions. We model total demand as $d_a(t)=\Bar{d}_a(t)+w_d(t)$, where $\Bar{d}_a$ represents known demand forecasts and $w_d$ is the uncertainty component assumed to be coming from set $W_d=\{w\in \mathbb{R}\mid w=e_dg,\left|\left|g\right|\right|_\infty\leq 1\}$ with $e_d\in \mathbb{R}$ being the scaling factor. The demand uncertainty set $W_d$ is identified similarly to the model uncertainty set $W_m$ using real demand data. Putting the demand model into the linear tank level model \eqref{eq:reducedModel}, we get
\begin{equation}
 \dot{{h}}(t)= Ah(t) + B_1u(t) + B_2 \Bar{d}_a(t)+ B_2w_d(t)+ w_m(t).
 \label{eq:modelWithDemand}
\end{equation}
To implement discrete control methods as described in Section~\ref{sec:control}, we discretize the continuous model with sampling time $\Delta t$ using fourth-order Runge-Kutta integration, where we assume $u$ and $d_a$ remain constant between sampling instances and obtain the following model
\begin{equation}
 h_{i+1}= A_dh_i + B_{d_1}u_i + B_{d_2} \Bar{d}_{a_i}+ B_{d_2}w_{d_i}+ B_{d_3}w_{m_i}.
 \label{eq:discrete2uncertainty}
\end{equation}
 Two uncertainties, $w_{d_i}\in W_d$ and $w_{m_i}\in W_m$, can be combined into a single disturbance, $w_i\in W$ as
\begin{equation}
 h_{i+1}= A_dh_i + B_{d_1}u_i + B_{d_2} \Bar{d}_{a_i} + w_i
 \label{eq:discrete}
\end{equation}
by appropriately defining the disturbance set $W$ to encompasses all the possible uncertainty scenarios from $w_{d_i}\in W_d$ and $w_{m_i}\in W_m$, we ensure $B_{d_2}W_d + B_{d_3}W_m \subseteq W$. In this work, we use a disturbance set $W$ of the form
\begin{equation}
\label{eq:uncertainty}
    W=\{w\in \mathbb{R}^n\mid w=Eg,\left|\left|g\right|\right|_\infty\leq 1\},
\end{equation}
  where $g$ is  the disturbance generator. The scaling matrix $E \in \mathbb{R}^{n\times l}$ is selected so that $B_{d_2}W_d + B_{d_3}W_m \subseteq W$ is satisfied.

 \section{Robust MPC}
 \label{sec:control}
This section presents a robust MPC method for pump scheduling in WDNs, adapted from \cite{Goulart2008EfficientRO}. While the core structure of the method remains consistent with the approach in \cite{Goulart2008EfficientRO}, we extend it by incorporating an additional term, $B_{d_2} \Bar{d}_{a_i}$, to account for the extra demand component in our system model \eqref{eq:discrete}.  Here, we reformulate the robust MPC approach outlined in \cite{Goulart2008EfficientRO} for \eqref{eq:discrete}. The section proceeds by posing the robust control problem, deriving a tractable formulation, and finally putting the problem in a more computationally efficient form. 

Assuming state measurements $h$ at each time step, the objective is to design a control law
$u=\mu_N(h)$
 that minimizes pump electricity costs while satisfying the operational constraints of the WDN under all disturbance $w$ sequences   over a finite horizon $N \in \mathbb{Z}_{\geq0}$.

The operational constraints considered in this paper include bounds on pump flow rates $u_i$, limited by pump capacities, and tank water levels $h_i$, restricted by tank capacities and minimum required storage volumes. Specifically, these constraints are expressed as
\begin{align}
\label{eq:ineqs}
    &Kh_i+Lu_i \leq b, \quad \forall i \in \mathbb{Z}_{\geq0}, \\
    &K=\begin{bmatrix}
        I_n & -I_n & 0 & 0
    \end{bmatrix}^T \in \mathbb{R}^{s\times n}, \nonumber \\ 
    &L=\begin{bmatrix} 0 & 0 & I_m & -I_m \end{bmatrix}^T \in \mathbb{R}^{s\times m}, \nonumber \\ 
    &b=\begin{bmatrix}
        h_{max}^T & h_{min}^T & u_{max}^T & 0
    \end{bmatrix}^T \in \mathbb{R}^{s} , \nonumber
\end{align}
where $u^{max} \in \mathbb{R}^m$ are the pump  upper flow limits and $h^{min},h^{max} \in \mathbb{R}^n$ are the limits on the tank water levels, and $s=2n+2m$ is the number of inequality constraints.


At each time $i$, we optimize over the input policies that are affine in the past disturbances, which can be written as
\begin{equation}
\label{eq:policy}    u_{j}=v_{j}+\sum_{k=i}^{j-1}{M_{j,k}w_k}, \quad \forall j \in \mathbb{Z}_{[i,i+N-1]},
\end{equation}
where $v_{j} \in \mathbb{R}^{m}$ and $M_{j,k} \in \mathbb{R}^{m\times n}$ are the optimization variables and $u_{j}$ is the input for time $j$.  The input sequence $\mathbf{u}= \begin{bmatrix} u^T_{i}\cdots u^T_{i+N-1} \end{bmatrix}^T \in \mathbb{R}^{m N}$ can be written in matrix form by defining the variable $\mathbf{v} \in \mathbb{R}^{mN}$ and the matrix $\mathbf{M} \in \mathbb{R}^{mN\times nN}$ as 
\begin{subequations}
\label{eq:Mandv}
\begin{align}
\mathbf{M}&=\left[\begin{matrix}0&\cdots&\cdots&0\\M_{i+1,i}&0&\cdots&0\\\vdots&\ddots&\ddots&\vdots\\M_{i+N-1,i}&\cdots&M_{i+N-1,i+N-2}&0\\\end{matrix}\right], \\
\mathbf{v}&=\left[v_{i}^T\cdots v_{i+N-1}^T\right]^T 
\end{align}
\end{subequations}
so that $\mathbf{u}=\mathbf{M}\mathbf{w}+\mathbf{v}$, where $\mathbf{w}=\begin{bmatrix}w^T_i\cdots w^T_{i+N-1} \end{bmatrix}^T \in \mathcal{W}=W \times \cdots \times W$. By further defining the matrix $J=I \otimes E$ so that $\mathbf{w}=J\mathbf{g}$, where $\mathbf{g}=\begin{bmatrix} g^T_i\cdots g^T_{i+N-1} \end{bmatrix}^T \in \mathbb{R}^{lN}$ is the disturbance generator sequence, the input sequence can also be written as $\mathbf{u}=\mathbf{M}J\mathbf{g}+\mathbf{v}$. The reason for choosing this policy is that, this class of input policies is equivalent to affine state feedback policies for linear systems \cite{GOULART2006523} and the set of admissible $(\mathbf{M},\mathbf{v})$ values are shown to be convex when the inequality constraints on the system are affine as in \eqref{eq:ineqs}. 

In this work, we use the electricity costs of the pumps for the disturbance-free input and state sequences as the cost function of the controller, which can be written as
\begin{align}
\label{eq:cost}
J(h,\mathbf{v},\mathbf{e})&=\sum_{j=i}^{i+N-1}e_jv_{j}^T(C\hat{h}_{j}+Dv_{j}-p_j^{in})
\end{align}
where   $\hat{h}_{i}=h$ and $\hat{h}_{j+1}= A_d\hat{h}_{j} + B_{d_1}v_{j} + B_{d_2} \Bar{d}_{a_j}, \: j=i,\cdots,i+N-1$ are the nominal states without the disturbance, $C\hat{h}_{j}+Dv_{j}$ corresponds to $p_j^{out}$ and $\mathbf{e}=\begin{bmatrix}e^T_i\cdots e^T_{i+N-1}\end{bmatrix}^T \in \mathbb{R}^{N}$ are the electricity prices. 

Putting everything together, the optimization problem for calculating the input policy at time instance $i$ can be written as
\begin{subequations}
\label{eq:optgeneral}
    \begin{align}
       &(\mathbf{M^{*}},\mathbf{v^{*}})= \argmin_{(\mathbf{M},\mathbf{v}) \in \Pi_N} J(h,\mathbf{v},\mathbf{e}),\\
        &\Pi_N = \left\{
     \mathbf{M}, \mathbf{v} \suchthat
     \begin{aligned}
       &\mathbf{(M},\mathbf{v}) \: \text{satisfies} \: \eqref{eq:Mandv}, h_{i}=h, \\ 
       &u_{j}=v_{j}+\sum_{k=i}^{j-1}{M_{j,k}w_k}, \\
        &h_{j+1}= A_dh_{j} + B_{d_1}u_{j}, \\
        &+ B_{d_2} \Bar{d}_{a_j}+w_j, \\ 
         &  Kh_{j}+Lu_{j} \leq b, K_hh_{i+N} \leq b_h, \\ & \forall w_j \in W, \: \forall j \in \mathbb{Z}_{[i, i+N-1]}
     \end{aligned}
  \right\} \label{eq:admFirst}
    \end{align}
\end{subequations}
where $K_h=\begin{bmatrix}I_n  & -I_n\end{bmatrix}^T \in \mathbb{R}^{2n\times n}$, $b_h=\begin{bmatrix}h_{max}^T & h_{min}^T\end{bmatrix}^T \in \mathbb{R}^{2n}$; $\mathbf{M^{i*}}$ and $\mathbf{v^{i*}}$ are the optimal values of relevant variables and $\Pi_N$ is the set of admissible policy parameters $(\mathbf{M},\mathbf{v})$. At each time instant, we like to solve this optimization problem and apply the resulting input policy to the system in a receding horizon fashion i.e. $\mu_N(h)=v_i^{*}$. However, in its current form, this problem is intractable because the constraints have to hold for $\forall w_j \in W$.  In \cite{Goulart2008EfficientRO}, an equivalent tractable problem is formed by redefining the admissible set $\Pi_N$. Although we have the additional demand term in the system model \eqref{eq:discrete}, the same steps can be used to reformulate the optimization problem \eqref{eq:optgeneral} to a tractable version.  In particular, the admissible set $\Pi_N$ can be equivalently written in a tractable form by designing the matrices  $\mathit{F} \in \mathbb{R}^{(sN+2n) \times mN}$, $\mathit{G} \in \mathbb{R}^{(sN+2n) \times nN}$, $\mathit{T} \in \mathbb{R}^{(sN+2n) \times n}$, $\mathbf{B_{2}} \in \mathbb{R}^{n(N+1) \times N}$, $\mathbf{K} \in \mathbb{R}^{(sN+2n) \times n(N+1)}$ and the vector $c \in \mathbb{R}^{sN+2n}$ as 
\begin{equation}
\label{eq:admissiblesstandart}
   \Pi_N = \left\{
     (\mathbf{M},\mathbf{v}) \suchthat
\begin{aligned}
&\mathbf{M},\mathbf{v} \: \text{satisfies} \: \eqref{eq:Mandv}, \exists \boldsymbol{\Lambda} \: s.t. \\
    &\mathit{F}\mathbf{v}+  \boldsymbol{\Lambda}\mathbf{1} \leq c- \mathbf{K}\mathbf{B_{2}}\mathbf{\Bar{d}_{a}}+\mathit{T}h, \\
    & -\boldsymbol{\Lambda} \leq \mathit{F}\mathbf{M}J+\mathit{G}J\leq \boldsymbol{\Lambda}
\end{aligned}
\right\}.
\end{equation}
where $\mathbf{\Bar{d}_{a}}=\begin{bmatrix}\Bar{d}_{a_i}^T\cdots \Bar{d}_{a_{i+N-1}}^T\end{bmatrix}^T \in \mathbb{R}^{N}$ are the demand predictions along the horizon and $\boldsymbol{\Lambda} \in \mathbb{R}^{(sN+2n)\times(lN)}$. With this formulation of $\Pi_N$, the optimization variables of \eqref{eq:optgeneral} becomes $\mathbf{M},\mathbf{v},\boldsymbol{\Lambda}$ and the total number of optimization variables becomes $mN$ in $\mathbf{v}$, $mnN(N-1)/2$ in $\mathbf{M}$, and $(sN+2n)lN$ in $\boldsymbol{\Lambda}$. For an interior point method, each iteration of problem \eqref{eq:optgeneral} would then take $\mathcal{O}(N^6)$ computational time. 

The exact expressions for matrices $\mathit{F}, \mathit{G},\mathit{T}$ and the vector $c$ can be found in \cite{Goulart2008EfficientRO}. The  matrices $\mathbf{K},\mathbf{B_{2}}$ are defined as
\begin{align}
\mathbf{B_{2}}&=\left[\begin{matrix}0&0&\cdots&0\\B_{d_2}&0&\cdots&0\\A_dB_{d_2}&B_{d_2}&\cdots&0\\ \vdots&\ddots&\ddots&\vdots\\A^{N-1}B_{d_2}&A^{N-2}B_{d_2}&\cdots&B_{d_2}\\\end{matrix}\right], \\
    \mathbf{K}&=\left[\begin{matrix}I_N\otimes K&0\\0 &I_n\\0&-I_n\end{matrix}\right]. 
\end{align}

\subsection{Computationally Efficient Formulation}
\label{sec:efficient}
In this section, we reformulate the constraints for the admissible set $\Pi_N$ and write them in a more computationally attractive form as done in \cite{Goulart2008EfficientRO}. We first introduce a new variable $\mathbf{U}=\mathbf{M}J$ such that $\mathbf{U} \in \mathbb{R}^{mN \times lN}$ has the same lower triangular structure as $\mathbf{M}$. The optimization problem \eqref{eq:optgeneral} becomes
\begin{subequations}
\label{eq:optUform}
    \begin{align}
&\min_{(\mathbf{U},\mathbf{v},\boldsymbol{\Lambda}) } J(h,\mathbf{v},\mathbf{e})\\
 &\mathbf{U}_{k,j}=0, \; \forall k \leq j, \\
         &\mathit{F}\mathbf{v}+  \boldsymbol{\Lambda}\mathbf{1} \leq c- \mathbf{K}\mathbf{B_{2}}\mathbf{\Bar{d}_{a}}+\mathit{T}h, \label{eq:lambdaNominal} \\
    & -\boldsymbol{\Lambda} \leq \mathit{F}\mathbf{U}+\mathit{G}J\leq \boldsymbol{\Lambda}. \label{eq:uConst}
    \end{align}
\end{subequations}
 Note that, with this formulation, the columns of $\mathbf{U}$ and $\boldsymbol{\Lambda}$ are decoupled in the constraints \eqref{eq:uConst}. Therefore, the constraints for different columns can be handled separately. In fact, each column of $\mathit{F}\mathbf{U}+\mathit{G}J$ can be written outputs of separate linear subsystems. Additionally, the constraints \eqref{eq:lambdaNominal} can be written by introducing a nominal subsystem.  These subsystems enable us to keep the state variables of the subsystems and write the constraints \eqref{eq:lambdaNominal}--\eqref{eq:uConst} in a sparse structure.

To introduce the nominal subsystem, we define the vector 
\begin{equation}
    \delta \mathbf{c}=[\delta c_{i}^T\cdots \delta c_{i+N}^T]^T=\boldsymbol{\Lambda}\mathbf{1} \label{eq:deltaDef}
\end{equation}
so that the constraint \eqref{eq:lambdaNominal}  becomes
\begin{equation}
    \mathbf{F}\mathbf{v}+  \delta \mathbf{c} \leq c- \mathbf{K}\mathbf{B_{2}}\mathbf{\Bar{d}_{a}}+\mathit{T}h. \label{eq:nominalMatrix} 
\end{equation}
It can be shown that the constraint \eqref{eq:nominalMatrix} can be written using the nominal states as  
\begin{subequations}
\label{eq:nominalConst}
    \begin{align}
    &\hat{h}_{i}=h, \label{eq:nominalinitial} \\
 &\hat{h}_{j+1}= A_d\hat{h}_{j} + B_{d_1}v_{j} + B_{d_2} \Bar{d}_{a_j}, \: \forall j \in \mathbb{Z}_{[i, i+N-1]},  \\
 &K\hat{h}_{j}+Lv_{j}+\delta c_{j} \leq b, \: \forall j \in \mathbb{Z}_{[i, i+N-1]}, \label{eq:nomSubsConst} \\
 & K_h\hat{h}_{j+N}  +\delta c_{i+N}\leq b_h, \label{eq:nomSubsConstTerm}
    \end{align}
\end{subequations}
which is in line with the classical nominal MPC setting but with an extra term $\delta c_{j}$ on the left-hand side of the constraints \eqref{eq:nomSubsConst} and \eqref{eq:nomSubsConstTerm}.

 Next, we deal with the constraints \eqref{eq:uConst} by introducing separate subsystems for each column of $\mathit{F}\mathbf{U}+\mathit{G}J$. We start with defining variables for columns for $\mathit{F}\mathbf{U}+\mathit{G}J$ and $\boldsymbol{\Lambda}$ as  
 \begin{subequations}
     \begin{align}
         &\mathbf{y^{p}}=(\mathbf{F}\mathbf{U}+GJ)e_p, \:  \label{eq:yp} \\
         &\delta \mathbf{c^{p}}=\boldsymbol{\Lambda}e_p, \:  \label{eq:deltapDef}
     \end{align}
 \end{subequations}
 where $e_p\in \mathbb{R}^{ln}$ is a unit vector whose $p$'th element is equal to $1$ and other elements are equal to $0$. With these definitions the inequality constraint \eqref{eq:uConst} can be written as
 \begin{equation}
 \label{eq:deltaIneq}
  -\delta \mathbf{c^{p}} \leq   \mathbf{y^{p}} \leq \delta \mathbf{c^{p}} , \:  \forall p \in \mathbb{Z}_{[1, lN]}.
 \end{equation}
It can be verified that $\mathbf{y^{p}}$ can be considered as the stacked values of $Kh_{j} + Lu_{j}$ of a system with $\mathbf{v}=0$, $h=0$, $\mathbf{\Bar{d}_{a}}=0$ and $\mathbf{g}=e_p$ as 
\begin{subequations}
\begin{align}
    & (u^p_{j}, h^p_{j},y^p_{j})=0, \: \forall j \in \mathbb{Z}_{[i, k]}, \label{eq:uhy=0}\\ 
    &h^p_{k+1}=E_{(z)}, \label{eq:hE}\\
    &h^p_{j+1}=A_dh^p_{j}+B_{d_1}u^p_{j} , \: \forall j \in \mathbb{Z}_{[k+1, i+N-1]}, \label{eq:hEdyn} \\
    &y^p_{j}=Kh^p_{j}+Lu^p_{j}, \: \forall j \in \mathbb{Z}_{[k+1, i+N-1]}, \label{eq:yout} \\
    &y^p_{i+N}=K_hh^p_{i+N}, \label{eq:yfinalout}
\end{align}
\end{subequations}
where $E_{(z)}$ is the zth column of $E$, $p=l(k-i)+z$ so that $k=\lfloor \frac{p-1}{l} \rfloor+i$, $z=1+ ((p-1)\: \text{mod} \: l)$ and  $\mathbf{y^{p}}=\begin{bmatrix}{y^p_{i}}^T\cdots {y^p_{i+N}}^T\end{bmatrix}^T$. By further defining $\delta \mathbf{c^{p}}$ as $\delta \mathbf{c^{p}}=\begin{bmatrix}{\delta c^p_{i}}^T\cdots {\delta c^p_{i+N}}^T\end{bmatrix}^T$, we can rewrite the inequalities \eqref{eq:deltaIneq} in the form 
\begin{equation}
\label{eq:deltasmall}
  -\delta c^p_{j}  \leq y^p_{j} \leq \delta c^p_{j},  \:  j \in \mathbb{Z}_{[i, i+N]}, \:  \forall p \in \mathbb{Z}_{[1, lN]}.
\end{equation}
Note that the inequalities for $j \leq k$ can be discarded since $y^p_{j}=0$ for $j \leq k$, (see \eqref{eq:uhy=0}), which means $\delta c^p_{j}$ are also $0$ because of \eqref{eq:deltasmall}. By defining
\begin{subequations}
    \begin{align}
       & \Bar{K}=\left[\begin{matrix} K \\ -K \end{matrix} \right], \Bar{L}=\left[\begin{matrix} L \\ -L \end{matrix} \right],\Bar{K_h}=\left[\begin{matrix} K_h \\ -K_h \end{matrix} \right], \\
       &H=\left[\begin{matrix} I_s \\ -I_s \end{matrix} \right],H_f=\left[\begin{matrix} I_{2n} \\ -I_{2n} \end{matrix} \right]
    \end{align}
\end{subequations}
the equations \eqref{eq:yout}, \eqref{eq:yfinalout} and the inequalities \eqref{eq:deltasmall} can be combined as
\begin{subequations}
\label{eq:deltapineq}
    \begin{align}
        &\Bar{K}h^p_{j}+\Bar{L}u^p_{j}+H\delta c^p_{j} \leq 0, \: \forall j \in \mathbb{Z}_{[k+1, i+N-1]}, \\
        &\Bar{K_h}h^p_{i+N}+H_f\delta c^p_{i+N} \leq 0.
    \end{align}
\end{subequations}
Note that by definitions of $\delta \mathbf{c}$ (see \eqref{eq:deltaDef}) and $\delta \mathbf{c^{p}}$ (see \eqref{eq:deltapDef}), it follows that
\begin{equation}
\label{eq:deltaRela}\textstyle
    \delta c_{i}^T= \sum_{p=1}^{lN}{\delta c^p_{j}} , \: \forall j \in \mathbb{Z}_{[i, i+N]}.
\end{equation}
\subsection{Complete problem}
We can now rewrite an optimization problem equivalent to \eqref{eq:optgeneral} as

    \begin{align}
    \label{eq:effProblem}
&\min_{\substack{\hat{h}_{i},\cdots,\hat{h}_{i+N},v_{i},\cdots,v_{i+N-1}\\ h^1_{i},\cdots,h^1_{i+N},u^1_{i},\cdots,u^1_{i+N},\delta c^1_{i},\cdots,\delta c^1_{i+N}\\ \cdots,\\ h^{lN}_{i},\cdots,h^{lN}_{i+N},u^{lN}_{i},\cdots,u^{lN}_{i+N},\delta c^{lN}_{i},\cdots,\delta c^{lN}_{i+N}  }} J(\hat{h}_{i},\mathbf{v},\mathbf{e}) \\
&\text{subject to \eqref{eq:nominalConst}, \eqref{eq:deltaRela}, \eqref{eq:uhy=0}-\eqref{eq:hEdyn} and \eqref{eq:deltapineq}. }\nonumber 
\end{align}
Note that the $\delta \mathbf{c}$ variables are not included as optimization variables, since they can be obtained from $\delta \mathbf{c^{p}}$ through \eqref{eq:deltaRela}. 

It was shown in \cite{Goulart2008EfficientRO} that a problem of this type can be put into a matrix form with a sparse structure
and can be solved with a per-iteration computational effort of $\mathcal{O}(N^3)$ using a primal-dual interior-point method for a convex quadratic cost function. Although the exact expressions for these matrices are not provided in this text, readers can find these expressions in \cite{Goulart2008EfficientRO} and in the accompanying code\footnote{The code is available at {\url{https://github.com/mirhanu/DisturbanceFeedback}. }}. Note that while the cost function \eqref{eq:cost} in this paper is quadratic, it is not necessarily convex.  However, even in the nonconvex case, the same per-iteration complexity of $\mathcal{O}(N^3)$ can be achieved using Sequential Quadratic Programming (SQP) or an interior point method with a positive definite Hessian approximation \cite{Wchter2006OnTI},\cite{nonconqp}. Another approach to solving the same type of robust optimization problem as in \eqref{eq:optgeneral} is presented in a recent work~\cite{leeman2024fastlevelsynthesisrobust}. While they do not specifically target water distribution networks, they offer an efficient solution with a computational complexity of $\mathcal{O}(N^2(n_x^3 + n_u^3))$ per iteration.

\section{Application}
\label{sec:application}
The presented method is implemented on an approximate model \eqref{eq:discrete} of the WDN of Randers,
a Danish city of approximately 64,000 people.  The network has 4549 nodes, 4905 links, and 8 pumping stations, of which we control two that supply three tanks. Since two tanks are connected by a large pipe with nearly equal levels, they are represented by a single variable, resulting in $n=2$ state variables and $m=2$ control variables. 

The constant matrices $A, B_1, B_2$, and model uncertainty set $W_m$ are identified using data from the EPANET model (not publicly available due to proprietary restrictions). as explained in Section \ref{sec:network}. The resulting matrices $A_d,B_{d_1},B_{d_2}$ for the discrete model \eqref{eq:discrete} and $E_m$ matrix defining the model uncertainty set $W_m$  are as follows 
\begin{align*}
     &A_d=\left[\begin{matrix} 0.9867 \quad    0.0134 \\ 0.0417 \quad   0.9577 \end{matrix} \right], \quad B_{d_1}=\left[\begin{matrix} 0.0013  \quad  0.0005 \\ 0.0008  \quad  0.0035 \end{matrix} \right], \\ &B_{d_2}=\left[\begin{matrix} -0.0012 \\ -0.0014 \end{matrix} \right], \quad E_m=\left[\begin{matrix} 0.054 \quad 0 \\ 0 \quad 0.083  \end{matrix} \right].
\end{align*}
The known part of the demand $\Bar{d}_a(t)$ is taken from the EPANET model of the system. As for the demand uncertainty set $W_d$, it is assumed that the bound on the absolute value of the demand uncertainty is 10 percent of the maximum known demand value.

\subsection{Simulation Results}
The network model \eqref{eq:discrete} is simulated with tank levels constrained between half-full (for emergency storage) and maximum levels (3 m for $h^1$, 2.8 m for $h^2$). The pump flow rates are bounded by 100 L/s, with a sampling time $\Delta t$ of 1 hour and a prediction horizon $N$ of 24 hours. The disturbance $w_i$ is calculated as $w_i=Eg_i$, where a disturbance generator $g_i$ is sampled for each time step. Two disturbance sets are tested: model uncertainty alone, and model uncertainty plus 10\% demand uncertainty.

In the simulations, the proposed disturbance feedback-based MPC (DFMPC) is compared with several other pump scheduling methods. One of these is a nominal MPC (NoMPC), for which the optimization problem is the same as the one given in \eqref{eq:optgeneral}, but with disturbances $w_i$ set to $0$. We also test the presented method against Constraint-Tightening MPCs (CTMPC) with tightened state constraints, where the state constraints are tightened with different amounts. Let $w^{\prime} \in W$ be the elementwise maximum possible disturbance. We use three CTMPCs: one (CTMPC1) with tightening of amount $w^{\prime}$, another (CTMPC1.5) with $1.5w^{\prime}$, and another (CTMPC2) with $2w^{\prime}$. Note that for all of these methods, soft state constraints with linear penalties on slack variables are used due to the possibility of constraint violations. 


To evaluate the methods, we simulated the network model using disturbance generators sampled from three sets representing different levels of difficulty: normal ($[-1,1]\times[-1,1]$), challenging ($[-1,-0.5]\times[0.5,1]$), and extreme ($\{-1\}\times\{1\}$). The challenging and extreme sets create opposing disturbances for the two tanks, making constraint satisfaction increasingly difficult. For each set, we generated one disturbance sequence and simulated the network using five different methods over 100 days, with computations performed in parallel 10-day blocks.

\begin{figure}[t]
        \includegraphics[width=5cm]{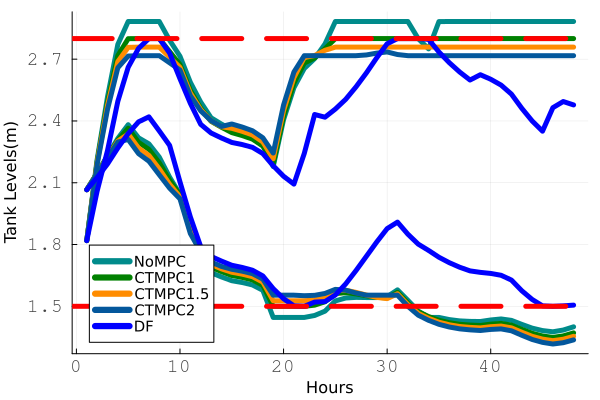}
        \centering
        \caption{The evolution of water levels in the two tanks over two days using various methods, with a disturbance generator $g_i=[-1 \; 1]$ for all $i$ considering only model uncertainties. Blue curves represent DFMPC while the horizontal red lines indicate the bounds on the tank levels. All methods except DFMPC result in constraint violations.}
        \label{fig:tank}
\end{figure}

In the first experiments considering only model uncertainties ($W=W_m$), Table \ref{table:model} shows the average daily costs and constraint violations over 100 days. DFMPC achieved the lowest cost for challenging and extreme sets, while for the normal set, its cost was only 3.2\% higher than the best method (CTMPC1). All methods except NoMPC satisfied constraints for normal and challenging sets. However, in the extreme set, only DFMPC maintained feasibility throughout the simulation, while other methods violated constraints extensively. As shown in Figure~\ref{fig:tank}, this occurred because DFMPC preemptively kept tank levels away from their bounds, while other methods allowed both tanks to approach their limits simultaneously, making constraint satisfaction impossible. These results show that DFMPC effectively maintains constraint satisfaction without significantly increasing costs when the pump capacity allows sufficient flexibility in flow adjustments.

\begin{table}
\caption{Simulation results considering only model uncertainty. The table shows average daily costs over 100 days and total state constraint violations for three disturbance sets, each tested with five methods.}
\label{table:model}
\resizebox{0.8\columnwidth}{!}{%
        \begin{tabular}{llll}\toprule
            \textbf{Disturbance generator set}&&\textbf{Average Daily Cost (€)} & \textbf{$\#$ of Constraint Violations} \\\midrule
            Normal ($[-1,1]\times[-1,1]$)  & DFMPC & 421.0 &  0 \\
                 & NoMPC & 387.2 & 391 \\
                  &\textbf{CTMPC1} &\textbf{407.9}  & \textbf{0} \\
                  &CTMPC1.5 &418.3  & 0 \\
                  &CTMPC2 &428.9  & 0 \\ \hline
            Challenging($[-1,-0.5]\times[0.5,1]$)  & \textbf{DFMPC} & \textbf{583.2} &  \textbf{0} \\
                 & NoMPC & 632.5 & 1160 \\
                  &CTMPC1 &664.2  & 0 \\
                  &CTMPC1.5 &681.5  & 0 \\
                  &CTMPC2 &699.5  & 0 \\ \hline
            Extreme($\{-1\}\times\{1\}$)  & \textbf{DFMPC} & \textbf{691.9} &  \textbf{0} \\
                 & NoMPC & 768.1 & 2266 \\
                  &CTMPC1 &765.3  & 2088\\
                  &CTMPC1.5 &762.7  & 2092 \\
                  &CTMPC2 & 759.5  & 2084 
            \\\bottomrule
        \end{tabular}}
\end{table} 


The second experiment considered both model and demand uncertainties (10\%  of maximum demand), using the same pump flow bounds but introducing soft state constraints due to occasional DFMPC infeasibility. As shown in Table~\ref{table:modelDemand10}, the cost of DFMPC, relative to the best performing method, increases to 8.5\% compared to 3.2\% in the first set of experiments, which means that a larger disturbance set had a greater impact on its cost performance compared to other methods, limiting its ability to follow optimal trajectories. It can be seen that, DFMPC maintained feasibility in challenging scenarios while other methods showed frequent violations, though all methods including DFMPC struggled in extreme cases. Compared to the first experiment, the pumps had less capacity to counteract disturbances in these experiments, reducing DFMPC’s cost efficiency and constraint satisfaction performance. Nevertheless, DFMPC still outperformed other methods in maintaining feasibility.

Computationally, DFMPC is more demanding than other methods due to its optimization over disturbance-dependent control policies. For a 24-hour horizon, it takes 4.11s per iteration compared to 0.24s for CTMPC1 on an Intel i5-1135G7 processor. While this is a drawback, the efficient formulation reduces per-iteration complexity from $\mathcal{O}(N^6)$ to $\mathcal{O}(N^3)$. Moreover, WDNs' slow dynamics allow hourly computation, making this practical for smaller networks. Additional plots illustrating the computational scaling with respect to the prediction horizon are available in the code repository.



\begin{table}[t]
\caption{Simulation results for various methods considering both model and 
 10 percent demand uncertainty.}
\label{table:modelDemand10}
\resizebox{0.8\columnwidth}{!}{%
        \begin{tabular}{llll}\toprule
            \textbf{Disturbance generator set}&&\textbf{Average Daily Cost (€)} & \textbf{$\#$ of Constraint Violations} \\\midrule
            Normal($[-1,1]\times[-1,1]$)  & DFMPC & 457.3 &  0 \\
                 & NoMPC & 395.2 & 378 \\
                  &\textbf{CTMPC1} &\textbf{421.5}  & \textbf{0} \\
                  &CTMPC1.5 &435.4  & 0 \\
                  &CTMPC2 &449.3  & 0 \\ \hline
            Challenging($[-1,-0.5]\times[0.5,1]$)  & \textbf{DFMPC} & \textbf{705.2} &  \textbf{17} \\
                 & NoMPC & 769.3 & 2214 \\
                  &CTMPC1 &769.0  & 1769 \\
                  &CTMPC1.5 &766.9  & 1723 \\
                  &CTMPC2 &763.4  & 1614 \\ \hline
            Extreme($\{-1\}\times\{1\}$)  & DFMPC & 755.3 &  2068 \\
                 & NoMPC & 764.9 & 2298 \\
                  &CTMPC1 &759.4 & 2240\\
                  &CTMPC1.5 &756.5  & 2226 \\
                  &CTMPC2 & 753.3  & 2230 
            \\\bottomrule
        \end{tabular}}
\end{table}

\section{Conclusion}
\label{sec:conclusion}
A robust MPC-based pump scheduling method with an efficient formulation of the robust optimization problem is presented for the optimal and reliable operation of WDNs. Testing against other MPC approaches on the Randers WDN model demonstrates that the proposed method performs best in terms of constraint satisfaction under high uncertainty when pump capacity is sufficient while maintaining competitive cost efficiency. However, when pump capacity is insufficient, all methods struggle to meet constraints, with the proposed method still performing the best in terms of constraint satisfaction but potentially at a higher cost. The DFMPC method requires more computational power than other methods. However, due to the slow dynamics of WDNs, the computation time may still be negligible for smaller systems, even with less powerful computers, while larger WDNs may require stronger computing resources.


\bibliographystyle{IEEEtran}
\bibliography{references}

\begin{thebibliography}{10}
\providecommand{\url}[1]{#1}
\csname url@samestyle\endcsname
\providecommand{\newblock}{\relax}
\providecommand{\bibinfo}[2]{#2}
\providecommand{\BIBentrySTDinterwordspacing}{\spaceskip=0pt\relax}
\providecommand{\BIBentryALTinterwordstretchfactor}{4}
\providecommand{\BIBentryALTinterwordspacing}{\spaceskip=\fontdimen2\font plus
\BIBentryALTinterwordstretchfactor\fontdimen3\font minus \fontdimen4\font\relax}
\providecommand{\BIBforeignlanguage}[2]{{%
\expandafter\ifx\csname l@#1\endcsname\relax
\typeout{** WARNING: IEEEtran.bst: No hyphenation pattern has been}%
\typeout{** loaded for the language `#1'. Using the pattern for}%
\typeout{** the default language instead.}%
\else
\language=\csname l@#1\endcsname
\fi
#2}}
\providecommand{\BIBdecl}{\relax}
\BIBdecl

\bibitem{epaEPANET}
``{E}{P}{A}{N}{E}{T} | {U}{S} {E}{P}{A} --- epa.gov,'' \url{https://www.epa.gov/water-research/epanet}, [Accessed 09-10-2024].

\bibitem{WSSEnergy}
N.~Sharif, H.~Haider, A.~Farahat, K.~Hewage, and R.~Sadiq, ``Water energy nexus for water distribution systems: A literature review,'' \emph{Environmental Reviews}, vol.~27, 03 2019.

\bibitem{MALAJETMAROVA2017209}
H.~Mala-Jetmarova, N.~Sultanova, and D.~Savic, ``Lost in optimisation of water distribution systems? {A} literature review of system operation,'' \emph{Environmental Modelling \& Software}, vol.~93, pp. 209--254, 2017.

\bibitem{Urkmez2024OptimizingPP}
M.~{\"U}rkmez, C.~S. Kalles{\o}e, J.~D. Bendtsen, and J.~Leth, ``Optimizing photovoltaic panel quantity for water distribution networks,'' \emph{ArXiv}, vol. abs/2412.15402, 2024.

\bibitem{Wang2018EconomicMP}
Y.~Wang, T.~Alamo, V.~Puig, and G.~Cembra{\~n}o, ``Economic model predictive control with nonlinear constraint relaxation for the operational management of water distribution networks,'' \emph{Energies}, vol.~11, p. 991, 2018.

\bibitem{chance1}
P.~Khatavkar and L.~W. Mays, ``Model for optimal operation of water distribution pumps with uncertain demand patterns,'' \emph{Water Resources Management}, vol.~31, pp. 3867--3880, 2017.

\bibitem{grosso}
J.~M. Grosso, P.~Velarde, C.~Ocampo-Martinez, J.~M. Maestre, and V.~Puig, ``Stochastic model predictive control approaches applied to drinking water networks,'' \emph{Optimal Control Applications and Methods}, vol.~38, no.~4, pp. 541--558, 2017.

\bibitem{PALLOTTINO20051031}
S.~Pallottino, G.~M. Sechi, and P.~Zuddas, ``A {DSS} for water resources management under uncertainty by scenario analysis,'' \emph{Environmental Modelling \& Software}, vol.~20, no.~8, pp. 1031--1042, 2005, methods of Uncertainty Treatment in Environmental Models.

\bibitem{9765702}
Y.~Guo, S.~Wang, A.~F. Taha, and T.~H. Summers, ``Optimal pump control for water distribution networks via data-based distributional robustness,'' \emph{IEEE Transactions on Control Systems Technology}, vol.~31, no.~1, pp. 114--129, 2023.

\bibitem{Goryashko2014RobustEC}
A.~Goryashko and A.~Nemirovski, ``Robust energy cost optimization of water distribution system with uncertain demand,'' \emph{Automation and Remote Control}, vol.~75, pp. 1754--1769, 2014.

\bibitem{perelman1}
G.~Perelman and A.~Ostfeld, ``Adjustable robust optimization for water distribution system operation under uncertainty,'' \emph{Water Resources Research}, vol.~59, no.~12, p. e2023WR035508, 2023.

\bibitem{7844498}
I.~Selek, E.~Ikonen, and C.~Hős, ``Tube-based robust {MPC} for pump scheduling in water distribution systems,'' in \emph{2016 IEEE International Conference on Systems, Man, and Cybernetics (SMC)}, 2016, pp. 001\,791--001\,796.

\bibitem{WANG20175202}
Y.~Wang, J.~Blesa, and V.~Puig, ``Robust periodic economic predictive control based on interval arithmetic for water distribution networks,'' \emph{IFAC-PapersOnLine}, vol.~50, no.~1, pp. 5202--5207, 2017, 20th IFAC World Congress.

\bibitem{Goulart2008EfficientRO}
P.~J. Goulart, E.~C. Kerrigan, and D.~Ralph, ``Efficient robust optimization for robust control with constraints,'' \emph{Mathematical Programming}, vol. 114, pp. 115--147, 2008.

\bibitem{GOULART2006523}
P.~J. Goulart, E.~C. Kerrigan, and J.~M. Maciejowski, ``Optimization over state feedback policies for robust control with constraints,'' \emph{Automatica}, vol.~42, no.~4, pp. 523--533, 2006.

\bibitem{Wchter2006OnTI}
A.~W{\"a}chter and L.~T. Biegler, ``On the implementation of an interior-point filter line-search algorithm for large-scale nonlinear programming,'' \emph{Mathematical Programming}, vol. 106, pp. 25--57, 2006.

\bibitem{nonconqp}
P.-A. Absil and A.~Tits, ``{Newton-KKT} interior-point methods for indefinite quadratic programming,'' \emph{Computational Optimization and Applications}, vol.~36, pp. 5--41, 07 2007.

\bibitem{leeman2024fastlevelsynthesisrobust}
A.~P. Leeman, J.~Köhler, F.~Messerer, A.~Lahr, M.~Diehl, and M.~N. Zeilinger, ``Fast system level synthesis: Robust model predictive control using {R}iccati recursions,'' 2024.

\end{thebibliography}
\end{document}